\documentclass[reprint,apstemplate]{revtex4-1}
\usepackage{bm,epsfig,graphics,amssymb,amsmath,subeqnarray,setspace,graphicx,amsthm,epstopdf,subfigure,color,physics,bm,url,hyperref}

\def\omegahat{\hat{\bm{\omega}}}
\def\Omegahat{\hat{\bm{\Omega}}}
\newcommand{\bhat}[1]{\bm{\hat{#1}}}

\newcommand{\Pexp}[1]{\text{Pexp}\left(#1\right)}
\def\Ai{\text{Ai}}
\def\Bi{\text{Bi}}
\pacs{47.63.-b, 47.63.Gd, 87.17.Jj, 87.23.Kg}

\begin{document}

\title{Self-buckling and self-writhing of semi-flexible microorganisms}
\author{Wilson Lough}
\email{wlough@wisc.edu}
\affiliation{Department of Physics, University of Wisconsin-Madison}
\author{Douglas B. Weibel}
\affiliation{Department of Biomedical Engineering, University of Wisconsin-Madison}
\author{Saverio E. Spagnolie}
\email{spagnolie@math.wisc.edu}
\affiliation{Department of Mathematics, University of Wisconsin-Madison}
\date{\today}

\begin{abstract}
The twisting and writhing of a cell body and associated mechanical stresses is an underappreciated constraint on microbial self-propulsion. Multi-flagellated bacteria can even buckle and writhe under their own activity as they swim through a viscous fluid. New equilibrium configurations and steady-state dynamics then emerge which depend on the organism's mechanical properties and on the oriented distribution of flagella along its surface. Modeling the cell body as a semi-flexible Kirchhoff rod and coupling the mechanics to a dynamically evolving flagellar orientation field, we derive the Euler-Poincar{\'e} equations governing dynamics of the system, and rationalize experimental observations of buckling and writhing of elongated swarmer cells of the bacterium {\it Proteus mirabilis}. A sequence of bifurcations is identified as the body is made more compliant, due to both buckling and torsional instabilities. These studies highlight a practical requirement for the stiffness of bacteria below which self-buckling occurs and cell motility becomes ineffective.
\end{abstract}
\maketitle

Motility introduces a number of demands on the mechanical construction of bacterial cells. Such constraints have been studied for motility organelles; slender flagella can buckle below a critical bending stiffness or above a critical motor torque \cite{vs12,jkdgr15}, and the same is true of the flexible flagellar hook \cite{ng18,zls21}. The shape and size of bacterial cells is influenced by numerous considerations \cite{Young06,myswnkfwwq12,wh17,slsvlj19}, including efficient motility in liquids \cite{agk04,sl10,shsh19}. However, motile bacterial cells are canonically presumed to be rod-shaped, non-deformable structures, and cell stiffness, a feature normally provided by cell wall composition \cite{tarhtscghw12,rboazmcwth18,agshr22} and turgor pressure \cite{rh18}, is typically overlooked in studies of motility. Cell wall stiffness regulation alters bacterial cell shape, influences motility, and enables bacteria to adapt and survive~\cite{alr16,aw17}.

Since the bending stiffness of an elongated body tends to be sensitive to its length, long cells can become highly deformed in complex or flowing environments. The length of {\it Proteus mirabilis} ({\it P. mirabilis}) cells, for instance, increases by up to 20-40x when they are in a swarming state \cite{Rather05}, and deformation in cell shape are visibly clear in a swarm \cite{cw09,aorlyjw19}. {\it P. mirabilis} swarmer cells have reduced cell stiffness compared to normal vegetative cells~\cite{aor19}. Gene deletion has also been used to artificially reduce cell stiffness \cite{tcapfsagw18}. But the nature and organization of any motility organelles is also important. A swarmer cell swims by rotating up to thousands of flagella which are distributed along its surface \cite{Kearns10,tc13}. The flagellar motion drives active, wavelike surface features more often used to describe ciliated organisms, which themselves are classically modeled as a continuum of active stress \cite{Blake72,bw77}. 

\begin{figure*}[htbp]
\includegraphics[width=\textwidth]{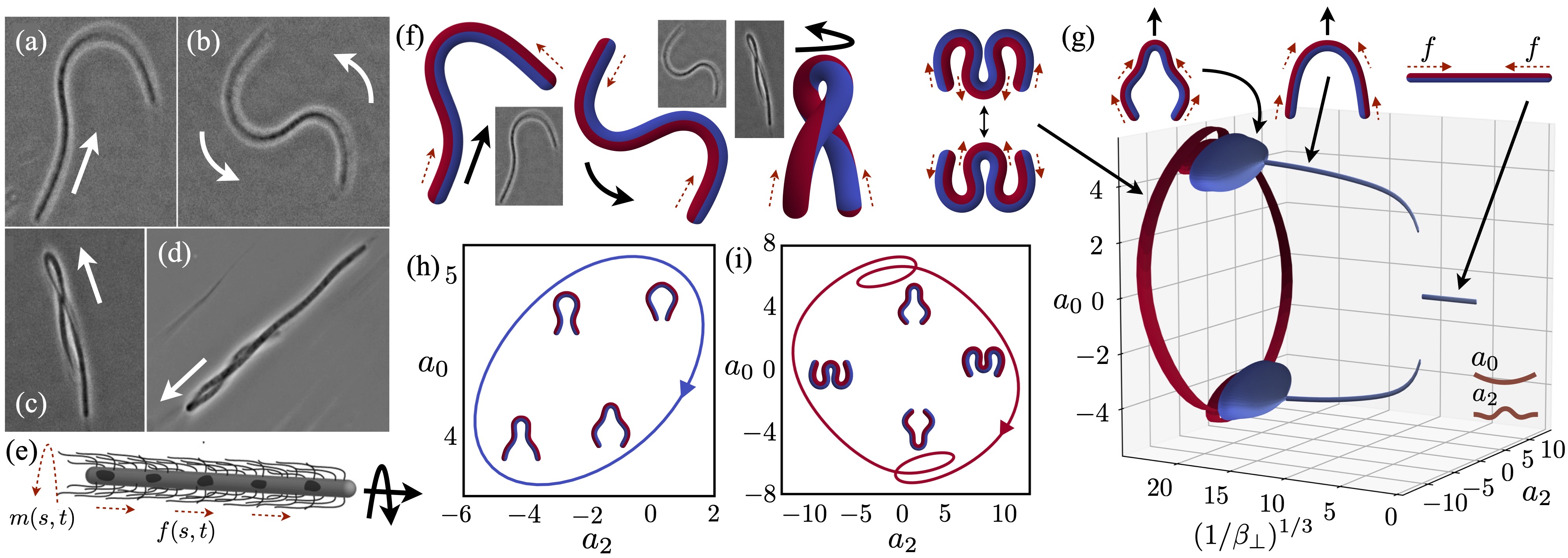}
\caption{(a-d) Swarmer {\it P. mirabilis} cells bend, rotate, and twist under their own flagellar activity (see Supplementary Movies M1-M4). Solid arrows indicate the direction of motion. (e) Flagellar stresses are modeled as a continuous force density $f(s,t)$ and proportional moment density $m=\chi f$ which drives and rotates the body through the fluid. (f) An active swimming Kirchhoff rod reproduces U-shaped swimming, S-shaped rotation, and twisted, rotating swimming states found in experiments. Dashed arrows indicate the direction of local flagellar forcing.  (g) Phase projection onto the first two even biharmonic modes for periodic symmetric dynamics with no active moment for a range of $\beta_\perp$. Bifurcations from straight filaments to swimming-U shapes, then to periodic waving-U dynamics, then to periodic flapping-W dynamics are observed as the bending stiffness is reduced. (h) A cross-section of the phase diagram in (g) with $\beta_\perp=1.3\times 10^{-4}$ (waving-U dynamics). (i) A cross-section of the phase diagram in (g) with $\beta_\perp=7.6\times 10^{-5}$ (flapping-W dynamics). }
\label{fig:Figure1}
\end{figure*}

A wild-type {\it P. mirabilis} cell is stiff and rod-shaped and swims along a straight trajectory, with its flagella oriented with their tips opposite the swimming direction (Fig.~\ref{fig:Figure1}e) \cite{Hoeniger65}. The fluid response to flagellar motion drives the body forward, and induces a rotational velocity along the long axis as dictated by the force- and torque-free nature of swimming in viscous fluids \cite{Lauga20}. Elongated swarmer cells, however, can express a wide range of intricate and stunning dynamics. Figure~\ref{fig:Figure1} shows {\it P. mirabilis} cells which have buckled under their own activity. The flagellar tips appear to be pointing away from the direction of {\it local} body motion, suggestive that their orientation depends upon local viscous stresses (Fig.~\ref{fig:Figure1}a-b; see Supplementary Movies M1-M4). Strongly three-dimensional configurations and dynamics are shown in Fig.~\ref{fig:Figure1}c, which includes a spinning motion about the direction of swimming. An even more highly deformed state with multiple self-crossings is shown in Fig.~\ref{fig:Figure1}d. 

Such active systems are particularly rich, as even passive slender bodies driven by external forces \cite{dlns19} or flows~\cite{ls15} continue to reveal new buckling behaviors~\cite{lmss13,pmclr15,ms15,lc18, cllcfrsl20,fngep22}. The shapes and dynamics of elongated {\it P. mirabilis} cells share many similarities with active or externally forced filaments which exhibit spontaneous symmetry breaking~\cite{jrglbka12,lgk18,spizvmgdd22}. The U- and S-shaped configurations in Fig.~\ref{fig:Figure1}a-b have been observed numerically in related systems in two dimensions~\cite{la15}, as have spiral-shaped configurations~\cite{ieg15}. The response of semi-flexible polymers to molecular-motor-driven stress has seen tremendous interest~\cite{weg17}, particularly in the context of cytoskeletal networks and interphase chromatin configurations~\cite{mk19,gg14,ssz18}. 

Flagellar propulsion, however, introduces additional features, for instance a competition between twist/bend elasticity and twist injection~\cite{wpg00,lp04,Powers10}, and a dynamic rearrangement of flagellar stress. It is plausible that the highly nonlinear twist-bend coupling~\cite{gpw98,gt97} responsible for the emergence of writhing instabilities~\cite{gt1997} and chiral configurations~\cite{ng99} in generic elastic filaments is also responsible for the configurations seen in Fig.~\ref{fig:Figure1}c-d.

In this paper we explore numerically and analytically a Kirchhoff rod model of a long, swimming cell which is driven by active forces and moments associated with flagellar activity. Body dynamics are described using the Euler-Poincar{\'e} formalism~\cite{gh09,br17,eghr11} which leverages the geometric structure of the Euclidean group $SE(3)$ and its Lie algebra $\mathfrak{se}(3)$ to seamlessly incorporate numerous kinematic constraints. The model reproduces both two- and three-dimensional configurations (Fig.~\ref{fig:Figure1}f) and predicts microorganism buckling and writhing under its own flagellar activity and viscous stress response. Bifurcations in the shapes and dynamics appear as the cell body is made more flexible, including buckling and torsional instabilities commonly observed in passive elastic systems, and new modes of motion are found upon the introduction of the active moment.

The cell is assumed to have length $L$ with uniform circular cross-section of diameter $a$. Aspect ratios $a/L$ of swarmers, typically on the order of $ 10^{-2}$ to $5\times10^{-2}$~\cite{Hoeniger65,ap19}, are sufficiently small that extensile and shear deformations are neglected~\cite{antmannonlinear}. Associated with each station of the filament in arclength $s$ and time $t$ is a
Euclidean transformation represented as a 4-by-4 matrix $\rho(s,t)=\left(\begin{matrix}
  Q & \vb{r}\\
  0 & 1
\end{matrix}
\right)$
which depends on the centerline position $\vb{r}$ and orthonormal material frame $Q=[\bm{q}_0|\bm{q}_1|\bm{q}_2]$. Body configurations are written as path ordered exponentials~\cite{sw89},
\begin{gather}
\label{eq:t_order}
    \rho(0,t) = \rho(0,0)\Pexp{\int_{0}^{t}\psi_t(0,\xi)\dd{\xi}},\\
    \label{eq:s_order}
    \rho(s,t) = \rho(0,t)\Pexp{\int_0^s\psi_s(\xi,t)\dd{\xi}},
\end{gather}
of $\mathfrak{se}(3)$-valued velocities $\psi_t=\rho^{-1}\partial_t\rho=\left(\begin{matrix}
    \omegahat & \vb{u}\\
    0 & 0
\end{matrix}\right)$ and deformations $\psi_s=\rho^{-1}\partial_s\rho=\left(\begin{matrix}
    \Omegahat & \vb{U}\\
    0 & 0
\end{matrix}\right)$, where we have defined the antisymmetric operators $\hat{\bm{\omega}}:=\bm{\omega}\times$ and $\Omegahat :=\bm{\Omega}\times$. Working in the local material frame, we formulate dynamics directly in terms linear and angular velocities, $\vb{u}=Q^{-1}\partial_t\bm{r}$ and $\bhat{\omega}=Q^{-1}\partial_tQ$, respectively, and their space-like analogues, the linear deformation $\vb{U}=Q^{-1}\partial_s \bm{r}$ and twist/curvature operator $\bhat{\Omega}=Q^{-1}\partial_s Q$. The well known compatibility relations~\cite{Powers10,antmannonlinear} for elastic rods are subsumed by the Euclidean structure equations~\cite{darling1994}
\begin{gather}
\label{eq:angular_structure}
    \partial_s\bm{\omega}-\partial_t\vb{\Omega} + \vb{\Omega}\times\bm{\omega}=\vb{0},
    \\
    \label{eq:linear_structure}
    \partial_s\vb{u}-\partial_t\vb{U} + \vb{\Omega}\times\vb{u}-\bm{\omega}\times\vb{U}=\vb{0},
\end{gather}
which ensure the integrability of the system $\rho^{-1}\dd{\rho}=\psi_s\dd{s}+\psi_t\dd{t}$. A principal advantage of this approach is that it naturally leads to numerical schemes which circumvent violations of inextensibility, unshearbility, and frame orthonormality, and do not require soft penalties or explicit parameterization of rotations by Euler angles or quaternions~\cite{im00,gf18,pc05,hl06,cg93}.

Viscous stresses, $\bm{\zeta}\cdot \bm{u}$ and $\zeta_r\bm{\omega}_\parallel=\left(\bm{U}\cdot\bm{\omega}\right)\bm{U}$, are characterized by the drag tensor, $\bm{\zeta}=\zeta_\parallel\vb{U}\vb{U}^T + \zeta_\perp(1-\vb{U}\vb{U}^T)$ with longitudinal $\zeta_\parallel$ and transverse $\zeta_\perp$ coefficients, and by a rotational drag coefficient $\zeta_r$~\cite{wpg00, Wada11}. Driving the system away from equilibrium are active stresses arising from a distribution of flagella, modeled here as a continuum providing an effective tangential force density $f\vb{U}$ and proportional moment density $m\vb{U}:=\chi f\vb{U}$  (Fig.~\ref{fig:Figure1}e). To account for the tendency of flagella to align with local flow, we consider $f$ to evolve according to
\begin{gather}
\label{eq:active}
    \tau_f \partial_t f = (F/L)\left[1-(f/F)^2\right]\vb{U}\cdot\vb{u}+D\partial_s^2f,
\end{gather}
with $\partial_sf(-L/2)=\partial_sf(L/2)=0$. The density tends toward a characteristic magnitude $F$ with a relaxation time $\tau_f\zeta_\parallel L/F$ depending on the dimensionless parameter $\tau_f$, and $D$ is a diffusion constant.

The internal energy of the body is given by $E = \int_{-L/2}^{L/2}\frac{1}{2} \bm{\Omega}\cdot B \bm{\Omega} + \bm{\Lambda}\cdot\bm{U}\dd{s}$, where $\bm{\Lambda}$ is a Lagrange multiplier which enforces inextensibility and unshearability,
\begin{gather}
\label{eq:inexten_unsher}
    \vb{q}_0=Q\vb{U}
\end{gather}
and $B=B_\parallel \vb{U}\vb{U}^T+B_\perp(1-\vb{U}\vb{U}^T)$ penalizes twisting and bending with moduli $B_\parallel$ and $B_\perp$, respectively~\cite{antmannonlinear}. Balancing structure preserving variations~\cite{gh09} of $E$ with active and viscous work gives the Euler-Poincar{\'e} equations,
\begin{gather}
\label{eq:force_balance}
\bm{\zeta}\cdot \bm{u}  = \partial_s\bm{\Lambda} + \bm{\Omega}\times\bm{\Lambda}+ f\bm{U},\\
\label{eq:moment_balance}
\zeta_r\bm{\omega}_\parallel = \partial_s\left(B\vb{\Omega}\right)+ \vb{\Omega}\times B\vb{\Omega}+\vb{U}\times\vb{\Lambda} + m\vb{U},
\end{gather}
which represent local force and moment balance~\cite{suppmat}. The kinematic relations~\eqref{eq:t_order}-\eqref{eq:linear_structure},\eqref{eq:inexten_unsher}, the flagellar evolution law \eqref{eq:active}, and the balance equations \eqref{eq:force_balance},\eqref{eq:moment_balance} form a closed system describing dynamics of the body and its flagellar distribution.

Using \eqref{eq:inexten_unsher}-\eqref{eq:moment_balance}, and the transverse part of \eqref{eq:linear_structure}, we solve for $\bm{U}$, $\bm{u}$, $\bm{\omega}$, and $\bm{\Lambda}_\perp:=(1-\vb{U}\vb{U}^T)\cdot\bm{\Lambda}$ in terms of the curvature $\bm{\Omega}$, active force density $f$ and tension, $\lambda:=\bm{\Lambda}\cdot \vb{U}$. These remaining variables evolve according to \eqref{eq:angular_structure}, \eqref{eq:active}, and the tangential component of \eqref{eq:linear_structure}, with force and moment-free endpoints requiring $\vb{\Omega}$, $\partial_s\vb{\Omega}_\perp$, and $\lambda$ to vanish at the boundaries~\cite{suppmat}. Equations \eqref{eq:t_order} and \eqref{eq:s_order} are used to locate and orient the body with respect to an inertial frame. Upon scaling by the length $L$, force density $F$, and stiffness $B_\perp$, the system is found to depend on five dimensionless groups: a relative bending modulus $\beta_\perp = B_\perp/(FL^3)$, twist modulus $\beta_\parallel = B_\parallel/(FL^3)$, translational drag ratio $\eta = \zeta_\perp/\zeta_\parallel$, rotational drag $\eta_r = \zeta_r/(\zeta_\parallel L^2)$, and scaled active moment $M = \chi/L$.  

The equations governing $(\bm{\Omega},f,\lambda)$ are discretized in space uniformly using second-order accurate central difference approximations, and advanced in time using a second-order implicit backward-differentiation scheme with a hybrid nonlinear solver applied at each timestep. Equations~\eqref{eq:t_order}-\eqref{eq:s_order} are solved using explicit second-order accurate Magnus integrators~\cite{in99,blanes2009magnus,suppmat}. Other approaches to this stiff numerical problem with different treatments of the hydrodynamics have recently been developed \cite{Lim10,gl12,olc13,lkol14,kl16,mspd22,wig23,gk22,ldr22,md22}. The parameters $(\eta, D, \tau_f)=(2, 10^{-3}, 10^{-2})$, timestep size $\Delta t =10^{-3}$, and spatial gridspacing $\Delta s=1/64$ are fixed for the duration unless otherwise stated.

In the case of no active moment, $M=0$, the body configuration is fully characterized by a single rotational strain, the (signed) centerline curvature $\kappa = \pm|\Omega_\perp|$. Restricting the shape evolution equations to two dimensions, the curvature and tension satisfy
\begin{multline}
\label{eq:nonlinear2d}
    \partial_{t}\kappa
    =
    -\frac{\beta_\perp}{\eta}\partial_{s}^{4}\kappa
+\frac{1}{\eta}\partial_{s}^{2}\left(
\kappa \lambda
\right)\\+
\partial_s\left(
\kappa\left[
\partial_s\lambda+f
\right]
\right)
+
\frac{\beta_\perp}{3}\partial_s^{2}\left(
 \kappa^{3}
\right),
\end{multline}
and
\begin{gather}
\label{eq:tension2d}
       \partial_{s}^{2}\lambda
   - \frac{\kappa^{2}}{\eta}\lambda
   =
   - \partial_{s}f
-
\frac{\beta_\perp}{2}\partial_s^2
\left(
 \kappa^2
\right)
-
\frac{\beta_\perp}{\eta}\kappa \partial_{s}^{2}\kappa.
\end{gather}

\begin{figure}[htbp]
\includegraphics[width=0.45\textwidth]{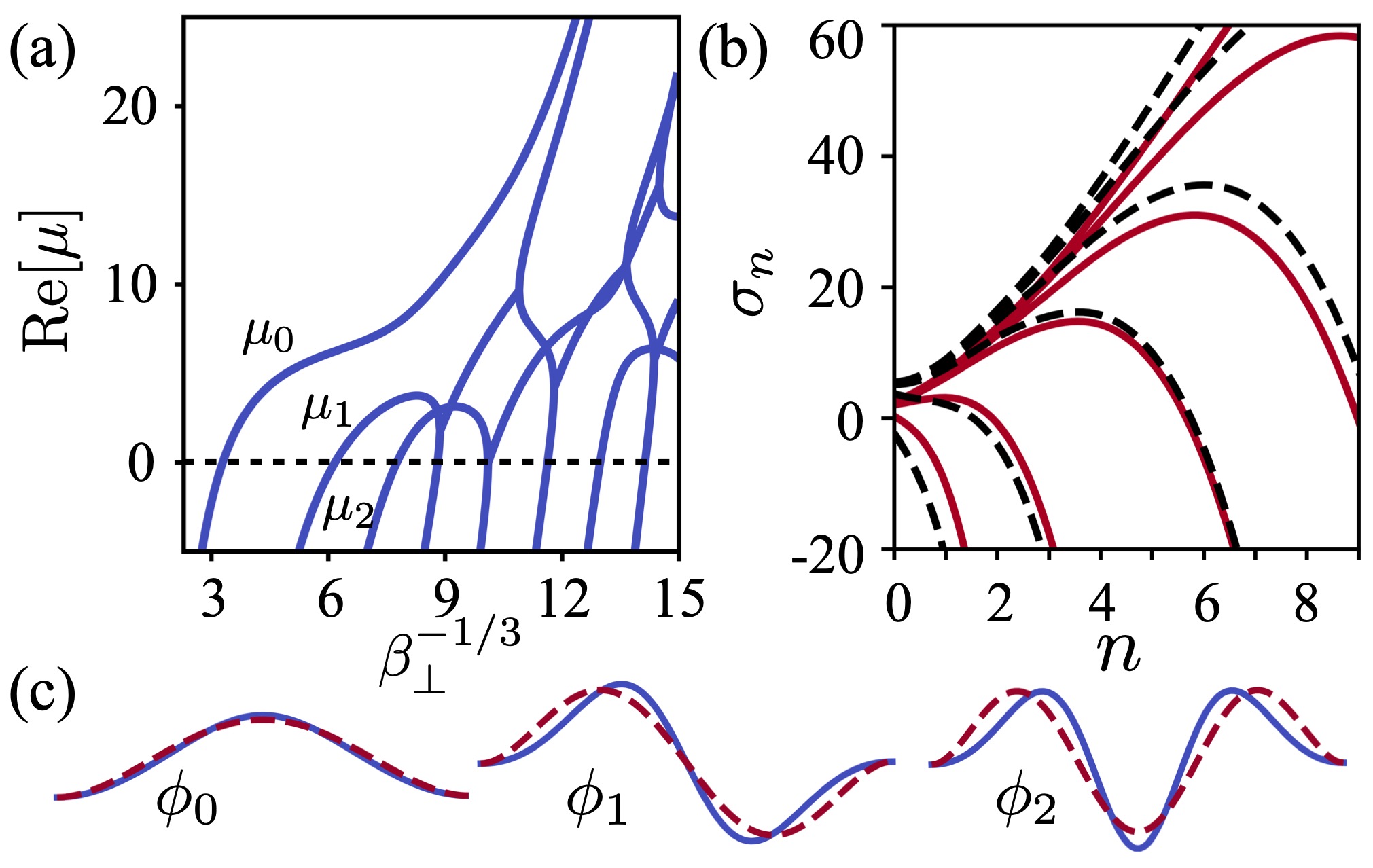}
\caption{(a) Dominant eigenvalues $\mu_n$ of the linearized curvature dynamics with no active moment show the emergence of multiple unstable modes at critical bending stiffnesses; the first at $\beta_\perp\approx 1.0\times 10^{-2}$.
(b) Growth rates of biharmonic eigenfunctions, $\phi_n(s)$, in the fully nonlinear system with $M=0$ (solid lines) and $M=0.01$ (dashed). (c) First three unstable modes of the linearized system (solid), and biharmonic eigenfunctions (dashed).
}
\label{fig:Figure2}
\end{figure}

The space of possible body motions is vast, so to begin we consider shapes which are symmetric about the body midpoint $s=0$ (and active forces which are odd). To describe the geometry it is convenient to use the eigenfunctions of $\partial_s^4$ satisfying force- and moment-free boundary conditions~\cite{wrog98} (the first three of which are shown in Fig.~\ref{fig:Figure2}c as dashed red curves). The curvature is decomposed as a sum $\kappa(s,t) = \sum_{k=0}^\infty a_{k}(t) \phi_{k}(s)$. Figure~\ref{fig:Figure1}g shows a phase portrait for the dynamics of the first two even biharmonic modes, $(a_0,a_2)$, for a range of bending stiffness $\beta_\perp$ with $M=0$. 

Phases in Fig.~\ref{fig:Figure1}g are identified by examining the long time behaviour of filaments initialized with a compressive active force density $f(s, 0)=-\tanh(10s)$. For $\beta_{\perp}>9.1\times 10^{-3}$, the stiff filament relaxes to a straight configuration, with the active stress eventually decaying due to diffusion via Eq.~\eqref{eq:active}. At approximately $\beta_{\perp}=9\times 10^{-3}$ there is a bifurcation to steady state U-shaped swimmers with a nonzero $a_0$ which dominates all other modes. Further decreases in stiffness lead to curvature oscillations (this cross-section of the phase diagram is shown in Fig.~\ref{fig:Figure1}h) and excitation of progressively higher modes. At approximately $\beta_\perp=6.8\times 10^{-5}$, another bifurcation is observed to unsteady, periodic flapping dynamics which involve even larger excursions in the phase plane (Fig.~\ref{fig:Figure1}i), and periodic changes in the swimming direction.

\begin{figure*}[thbp]
\includegraphics[width=\textwidth]{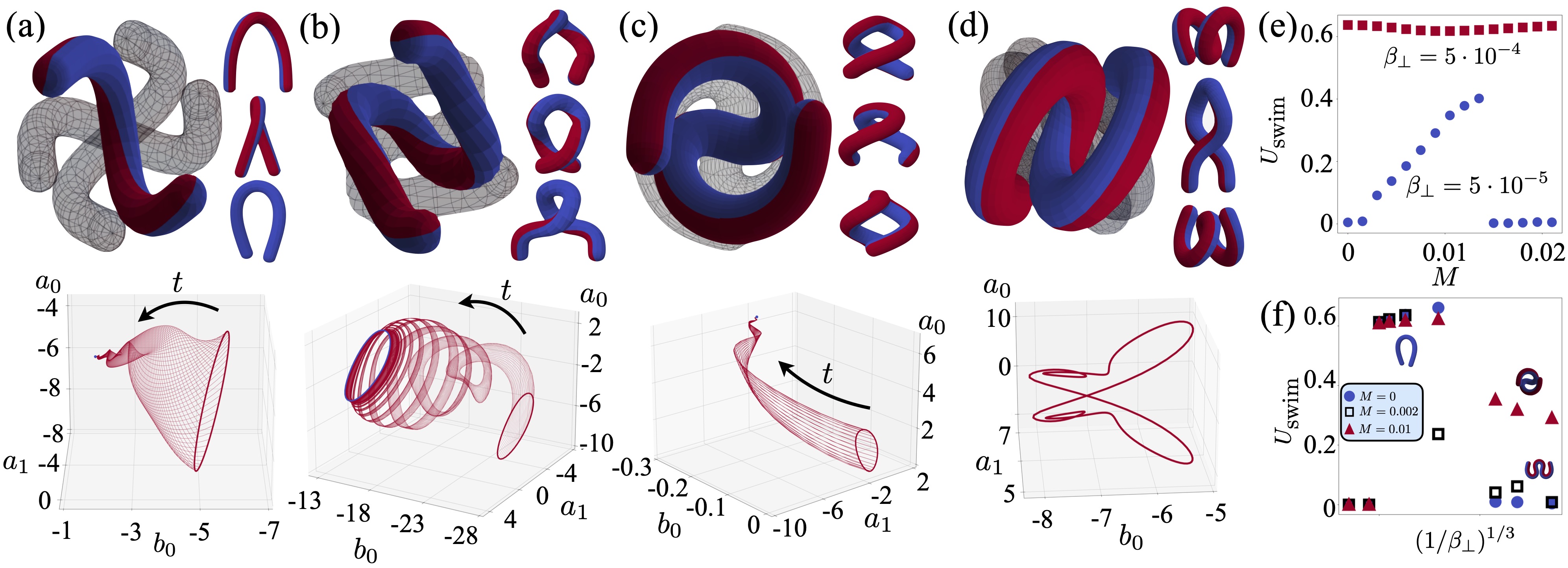}
\caption{(a-d) New attractors emerge upon the addition of an active moment (with body symmetry assumed). Top - snapshots of body configurations over a half-period, bottom - trajectories of the first two bending ($a_0, a_1$) and the first twisting ($b_0$) mode coefficients suggests convergence to a fixed shape. (a)
$(\beta_\parallel,\beta_\perp, M)=(1\times 10^{-4},2.5\times 10^{-4},6\times 10^{-3})$: a U-shaped swimmer with a twist. (b) 
$(\beta_\parallel,\beta_\perp, M)=(1\times 10^{-5},8\times 10^{-5},1.35\times 10^{-2})$:
reduced stiffness and increased active moment introduces a limit cycle corresponding to waving while twisting (see Supplementary Movie M5). (c) $(\beta_\parallel,\beta_\perp, M)=(1\times 10^{-3},5.45\times 10^{-5}, 1\times 10^{-2})$: convergence to a new twisted-S shape. (d) Periodic flapping appears with further increases in $M$ (see Supplementary Movie M6). (e) Swimming speed across a range of active moments for $\beta_\parallel=2.5\times 10^{-4}$, for a stiffer ($\beta_\perp = 5\times 10^{-4}$, squares) and softer ($\beta_\perp=5\times 10^{-5}$, circles) body. (f) Swimming speed across a range of bending stiffness for $\beta_\parallel=2.5\times 10^{-4}$ and active moments $M=0$ (circles), $M=0.002$ (squares), and $M=0.01$ (triangles).}
\label{fig:Figure3}
\end{figure*}

Susceptibility to buckling can be understood by exploring linear instabilities of a nearly straight body with a symmetric compressive force. Linearization of Eqs.~\eqref{eq:nonlinear2d}-\eqref{eq:tension2d} yields an eigenvalue problem, $\mathcal{L}[\kappa]=\mu\kappa$, where
\begin{gather}
\label{eq:linear2d}
   \mathcal{L}[\kappa]
=
-\frac{\beta_\perp}{\eta}\partial_{s}^{4}\kappa
-\frac{1}{\eta}\partial_{s}^{2}\left(\kappa\int_{-L/2}^{s}f\dd{s}
\right).
\end{gather}
Figure~\ref{fig:Figure2}a shows the (real part) of dominant eigenvalues $\mu_n$ of Eq.~\eqref{eq:linear2d} for a range of stiffness $\beta_\perp$. The unstable modes of the linear system are illustrated in Fig.~\ref{fig:Figure2}c by solid blue curves, along with the biharmonic eigenfunctions for comparison. Growth rates $\sigma_n=\left.\partial_t a_n(t)\right|_{t=0}/a_n(0)$, computed using the fully nonlinear system, Eqs.~\eqref{eq:nonlinear2d}-\eqref{eq:tension2d}, with $a_n(0)=10^{-3}$ are shown in Fig.~\ref{fig:Figure2}b. 

To identify the critical values of $\beta_\perp$ where different spatial modes become unstable, we set $\mu=0$ and seek solutions to $\mathcal{L}[\kappa]=0$. For a piecewise constant force density, $f(s)=1-2H(s)$, where $H(s)$ the Heaviside step function, critical values for even modes using Eq.~\eqref{eq:linear2d} are given by the solutions of
\begin{gather}
\label{eq:crit_even}
     \Bi'(\xi)\int_0^{\xi}\Ai(x)\dd{x}-\Ai'(\xi)\int_0^{\xi}\Bi(x)\dd{x}=0,
\end{gather}
where $\Ai$ and $\Bi$ are the Airy functions of the first and second kind, and $\xi=-\beta_\perp^{-1/3}/2$~\cite{suppmat}. For odd modes they are given by solutions of
\begin{gather}
-\xi\Bi(\xi)\int_0^{\xi}\Ai(x)\dd{x}
     +\xi\Ai(\xi)\int_0^{\xi}\Bi(x)\dd{x}
     \nonumber\\
     +\Bi'(0)\Ai(\xi)-\Ai'(0)\Bi(\xi)-1/\pi
     =0.\label{eq:crit_odd}
\end{gather}
When compared to the first ten critical stiffnesses in the fully nonlinear dynamics with regularized active force density, predictions of Eqs.~\eqref{eq:crit_even}-\eqref{eq:crit_odd} were found to differ by $0.5\%$-$15\%$. The first bending stiffness below which the filament becomes unstable from the full system is $\beta_\perp=1.0 \times 10^{-2}$, whereas the linearized dynamics predict $\beta_\perp=1.1 \times 10^{-2}$.

We turn now to the fully three-dimensional system, including the active moment contribution due to flagellar chirality ($M\neq 0$). Numerous dynamical regimes appear as the result of rotational forcing (see supplementary Movie M7), with transitions between newly emergent phases brought about by variations in any one of the twisting stiffness, $\beta_\parallel$, bending stiffness, $\beta_\perp$, or active moment, $M$. 

As with the 2D system, we seek a reduced order phase space in which to study these bifurcations. To this end, we consider systems initialized with the twist $\Omega_0$ and curvature $\Omega_2$ even about the midpoint, and the curvature $\Omega_1$ odd. This is equivalent to the system possessing a $\pi$-rotational symmetry, and, provided the initial active stress distributions are odd functions of $s$ about the midpoint, this symmetry is conserved. Taking advantage of their conserved parity, the twist and curvatures may be decomposed into sums of harmonic $\gamma_{2k}$ and biharmonic $\{\phi_{2k},\phi_{2k+1}\}$ functions satisfying appropriate parity and boundary and conditions: $\Omega_0(s,t)=\sum_{k}b_{2k}(t)\gamma_{2k}(s)$, $\Omega_1(s,t)=\sum_{k}a^{(1)}_{2k+1}(t)\phi_{2k+1}(s)$, $\Omega_2(s,t)=\sum_{k}a^{(2)}_{2k}(t)\phi_{2k}(s)$.

Figure~\ref{fig:Figure3}a-d shows characteristic shapes of four observed phases (top), as well as corresponding phase space trajectories of $(b_0,a_1,a_0):=(b_0, a_1^{(1)}, a_0^{(2)})$ appearing in the decomposition of twist/curvatures for a range of initial conditions (bottom). For $\beta_\parallel<2.5\times 10^{-4}$, $\beta_\perp>7.5\times 10^{-4}$, and $M>1.4\times 10^{-2}$ the body adopts a straight configuration. A bifurcation to a twisted-U shape appears upon increasing $\beta_\parallel$, decreasing $\beta_\perp$, or decreasing $M$ (Fig.~\ref{fig:Figure3}a). With $M<1.8\times 10^{-2}$, the twisted-$U$ phase persists as $\beta_\perp$ is decreased until approximately $\beta_\perp=2\times 10^{-4}$, at which point the system develops periodic oscillations (Fig.~\ref{fig:Figure3}b, see Supplementary Movie M5). Again with $M<1.8\times 10^{-2}$, new S-shaped equilibria emerge for $\beta_\perp <1\times 10^{-4}$ (Fig.~\ref{fig:Figure3}c). A fourth phase appears for $M>1.8\times 10^{2}$ and $\beta_\perp<2.5\times 10^{-4}$ with twist-curvature oscillations accompanied by periodic changes in swimming direction (Fig.~\ref{fig:Figure3}d, see Supplementary Movie M6).

Transitions between phases can lead to wide variations in swimming trajectories, and in the swimming speed, defined as the magnitude of the average velocity of the body's midpoint in the lab frame, $U_{swim}(T)=\left|\int_0^{T} Q(0,t)\vb{u}(0,t)\dd{t}\right|/T$. The complicated relationship between bend and twist is further illustrated by the nonmonotonic, and discontinuous, changes in swimming speed that arise due to variations in bending stiffness $\beta_\perp$ and active moment $M$. Figure~\ref{fig:Figure3}e shows the swimming speed as a function of the active moment for two different bending stiffnesses. For the stiffer body the active moment induces waving (from Fig.~\ref{fig:Figure3}a to Fig.~\ref{fig:Figure3}b, see Supplementary Movie M5) but the swimming speed remains roughly unchanged. For the softer body, however, which at $M=0$ is in the dramatic flapping-W state in two dimensions (Fig.~\ref{fig:Figure1}i), the introduction of the active moment can stabilize the shape in three dimensions and result in a ballistic trajectory (Fig.~\ref{fig:Figure3}c). Further increases in $M$, however, then trigger another phase transition to the three-dimensional flapping dynamics of Fig.~\ref{fig:Figure3}d (see Supplementary Movie M6), resulting in average speeds (but not instantaneous speeds) tending to zero. A different view is offered by Fig.~\ref{fig:Figure3}f, which shows the swimming speed across a range of bending stiffnesses for three different active moments. A sufficiently large active moment can delay the onset of flapping dynamics, and thereby stabilize swimming trajectories over a larger range of stiffnesses.

At the lower bending stiffness typical of swarmer cells, rotational forcing introduces a dynamical twist-bend instability. As shown in Fig.~\ref{fig:Figure2}b as dashed lines for $M=0.01$, the presence of an active moment can decrease the force required to excite higher unstable modes. As described in relation to Fig.~\ref{fig:Figure3}e above, this allows the system to access new energetically favorable out-of-plane equilibria similar to the `locked curvature' configurations observed in model cilia~\cite{lgk18,mk19}. 

Though not explored in detail here, both of the low stiffness configurations shown in Fig.~\ref{fig:Figure3}c,d are generically unstable with respect to asymmetric perturbations, which lead to non-periodic dynamics and trajectories which depend sensitively upon the bending stiffness (see Supplementary Movie M8). The self-contact evident in Fig.~\ref{fig:Figure1}d, and self intersections observed at low bending stiffness in the model, suggest that steric interactions are important for stabilizing body configurations of longer swarmer cells. Confinement by neighboring cells in bacterial swarms may play a similar role. 

Comparison of our results with experimental observations shows many behaviours of {\it P. mirabilis} swarmer cells are captured by the active Kirchhoff rod model. The relative bending stiffness $\beta_\perp=B_\perp/FL^3$, relating the flagellar stress to the cell's material and geometric properties, is seen to play an outsized role. Our analysis reveals a minimal value, approximately $\beta_\perp= 1.01\times 10^{-2}$, required of a cell below which its motility is severely hampered by self-buckling. For {\it P. mirabilis} swarmer cells, this corresponds to a critical bending stiffness of $B_\perp=3\times 10^{-23}\text{Nm}^2$, approximately one order of magnitude lower than the experimentally determined stiffness of typical cells~\cite{ap19}. That the discrepancy is not multiple orders of magnitude suggests that cells may actively maintain mechanical properties to prevent buckling during motility. This is a fascinating observation that may offer insight relevant to the evolutionary development of motility, bacterial adaptation and survival, and potential mechanically-motivated medical interventions. Additional experimental measures of twisting moduli of real biological cells, and more detailed treatment of the hugely complex, flagellated surface, would be needed to further probe this conjecture.

This work was supported by NSF Grant No. DMS-1661900.

\bibliographystyle{apsrev4-1}
\bibliography{BigBib}

\end{document}


\title{Self-buckling and self-writhing of semi-flexible microorganisms - Supplementary material}
\author{Wilson Lough, Douglas B. Weibel, and Saverio E. Spagnolie}
\date{\today}

\maketitle


\section{Kinematics}
\label{sec:kinematics}

\subsection{SE$(3)$ and $\mathfrak{se}(3)$}
\label{sec:se3} 
Kinematics and dynamics of the body are described using the geometric structure of the Euclidean group $SE(3)$ and its Lie algebra $\mathfrak{se}(3)$.
Euclidean transformations are represented by matrices 
\begin{gather}
    \left(
    \begin{matrix}
        q_{0x}&q_{1x}&q_{2x}&x\\
        q_{0y}&q_{1y}&q_{2y}&y\\
        q_{0z}&q_{1z}&q_{2z}&z\\
        0&0&0&1
    \end{matrix}
    \right),
\end{gather}
where $Q=\left(
    \begin{matrix}
        q_{0x}&q_{1x}&q_{2x}\\
        q_{0y}&q_{1y}&q_{2y}\\
        q_{0z}&q_{1z}&q_{2z}
    \end{matrix}
    \right)
    $ is an orthogonal matrix whose columns are interpreted as material frame vectors and $\bm{r} = \left(
    \begin{matrix}
        x\\
        y\\
        z
    \end{matrix}
    \right)$ is interpreted as the centerline position vector.
By a slight abuse of notation, we write these 4-by-4 $SE(3)$ matrices as $\SEmat{Q}{\bm{r}}$. The exponential map for $SE(3)$ can be expressed in closed form as
\begin{gather}
\label{eq:se3_exp}
     \exp
    \left(
    \begin{matrix}
        \bhat{\alpha} & \bm{b}\\
        0 & 0
    \end{matrix}
    \right)
    =
    \left(
    \begin{matrix}
        e^{\bhat{\alpha}} & A\bm{b}\\
        0 & 1
    \end{matrix}
    \right),
\end{gather}
where $e^{\bhat{\alpha}} = 1+\displaystyle\frac{\sin\alpha}{\alpha}\bhat{\alpha}
    +
    \frac{1-\cos\alpha}{\alpha^2}\bhat{\alpha}^2$, $ A = 1+\displaystyle\frac{1-\cos\alpha}{\alpha^2}\bhat{\alpha}
    +
    \frac{\alpha-\sin\alpha}{\alpha^3}\bhat{\alpha}^2$, and $\alpha = |\bm{\alpha}|$.

\subsection{Structure equations}
\label{sec:structure}
Deformation and velocity fields $(\psi_s, \psi_t)$ are components of an $\mathfrak{se}(3)$-valued one-form,
\begin{gather}
    \psi=\psi_s\dd{s}+\psi_t\dd{t}=\rho^{-1}\dd{\rho},
\end{gather}
where $\dd{\rho}=\partial_s\rho\dd{s}+\partial_t\rho\dd{t}$ is the exterior derivative of $\rho$, and $(\dd{s},\dd{t})$ are 1-forms dual to the coordinate basis $(\partial_s,\partial_t)$. Repeated application of the exterior derivative gives
\begin{gather}
    0=\dd[2]{\rho}=\rho\dd{\psi}+\dd{\rho}\wedge\psi,
\end{gather}
where $\wedge$ is the wedge product for matrix-valued differential forms~\cite{darling1994}. Multiplying by $\rho^{-1}$ and using $\psi=\rho^{-1}\dd{\rho}$ 
results in the Maurer-Cartan equation,
\begin{gather}
    \label{eq:structure_equation}
    \dd{\psi}+\psi\wedge\psi=0.
\end{gather}
This gives the commutation relation,
\begin{gather}
    \label{eq:st_commutator}
    \left(\dd{\psi}+\psi\wedge\psi\right)(\partial_s, \partial_t) = \partial_s\psi_t-\partial_t\psi_s+\comm{\psi_s}{\psi_t}=0,
\end{gather}
where $\comm{\psi_s}{\psi_t}:=\psi_s\psi_t-\psi_t\psi_s$ is the matrix commutator. Expressing \eqref{eq:st_commutator} in terms of $(\bhat{\Omega}, \bm{U}, \bhat{\omega},\bm{u})$, and using the fact that $\widehat{\bm{a}\times\bm{b}}=\comm{\bhat{a}}{\bhat{b}}$, for any $\bm{a},\bm{b}\in\mathbb{R}^3$, gives the structure equations found in the main text.

\subsection{Shape and spatial orientation}
\label{sec:shape_orientation}
We decompose $\rho(s,t)=\gamma(t)\sigma(s,t)$ into a transformation, $\sigma(s,t)=\rho(0,t)^{-1}\rho(s,t)$, describing the filament's shape relative to $s=0$, and a transformation, $\gamma(t)=\rho(0,t)$, locating and orienting the body in space. The state can then be recovered from $\psi$ by solving a pair of initial value problems,
\begin{gather}
  \label{eq:rho_t_ivp}
  \partial_t\gamma(t)=\gamma(t)\psi_t(0,t),
  \\
  \label{eq:rho_s_ivp}
  \partial_s\sigma(s,t)=\sigma(s,t)\psi_s(s,t),
\end{gather}
with initial conditions given by $\gamma(0)=\rho(0,0)$ and $\sigma(0,t)=1$. Solutions,
\begin{gather}
    \gamma(t) = \rho(0,0)\Pexp_0^t\psi_t(0,\xi)\dd{\xi},
    \\
    \sigma(s,t) = \Pexp_0^s\psi_s(\xi,t)\dd{\xi},
\end{gather}
are given in terms of the \emph{path-ordered exponential}~\cite{sw89}. If $\theta=\theta(\xi)$  is a curve through $\mathfrak{se}(3)$, the path-ordered exponential, or product integral, is defined by
\begin{gather}
  \Pexp_{a}^{b}\theta(\xi)\dd{\xi}=\prod_{a}^{b}e^{\theta(\xi)\dd{\xi}} = \lim_{n\rightarrow \infty} e^{\theta(\xi_0)\Delta\xi}e^{\theta(\xi_1)\Delta\xi} \ldots e^{\theta(\xi_{n})\Delta\xi},
\end{gather}
where $\Delta\xi=(b-a)/n$ and $\xi_k=a+k\Delta \xi$. 

\section{Dynamics}
\subsection{Euler-Poincar{\'e} variational principle \& balance equations}
\label{sec:ep_equations}
The variation $\delta\rho:=\left.\partial_\xi\tilde{\rho}(\xi)\right|_{\xi=0}$ is given by the derivative of a field, $\tilde{\rho}=\tilde{\rho}(s,t,\xi)$, which satisfies $\tilde{\rho}(s,t,0)=\rho(s,t)$~\cite{lrh89}. The associated 1-form $\tilde{\psi}=\tilde{\psi}_s\dd{s}+\tilde{\psi}_t\dd{t}+\tilde{\psi}_\xi\dd{\xi}:=\tilde{\rho}^{-1}\dd{\tilde{\rho}}$ now includes a $\xi$-component which acts as the generator of variations $\delta\rho = \left.\rho\tilde{\psi_\xi}\right|_{\xi=0}:=\rho\psi_\xi$. It follows directly from the definitions of $(\psi,\tilde{\psi},\tilde{\rho})$ that $\tilde{\psi}_s(s,t,0)=\psi_s(s,t)$, so,
 the variation $\delta\psi_s(s,t)=\left.\partial_\xi\tilde{\psi}_s(s,t,\xi)\right|_{\xi=0}$ is constrained by the Maurer-Cartan equation,
\begin{gather}
    \label{eq:variation_commutator}
    \left.\left(\dd{\tilde{\psi}+\tilde{\psi}\wedge\tilde{\psi}}\right)(\partial_s,\partial_\xi)\right|_{\xi=0}=\delta\psi_s-\partial_s\psi_\xi + \comm{\psi_s}{\psi_\xi}=0.
\end{gather}

We define the inner product for $\mathfrak{se}(3)$-valued matrices $\theta=\left(\begin{matrix}
    \bhat{\omega}_\theta & \bm{u}_\theta\\
    0&0
\end{matrix}\right)$ and $\phi=\left(\begin{matrix}
    \bhat{\omega}_\phi & \bm{u}_\phi\\
    0&0
\end{matrix}\right)$ by $ \langle\theta,\phi\rangle := \bm{\omega}_\theta\cdot\bm{\omega}_\phi+\bm{u}_\theta\cdot\bm{u}_\phi$ and the \emph{coadjoint operator}, $\coad{\psi_s}$, by $\langle \coad{\psi_s}\theta,\phi\rangle:=\langle\theta,\comm{\psi_s}{\phi}\rangle$. In terms of angular and linear components, we find $\coad{\psi_s}\phi
=
-
\left(\begin{matrix}
    \widehat{\bm{\Omega}\times\bm{\omega}_\theta} + \widehat{\bm{U}\times\bm{u}_\theta}& \bm{\Omega}\times\bm{u}_\theta\\
    0&0
\end{matrix}\right)$. With $\mathcal{E}=\displaystyle\frac{1}{2}\bm{\Omega}\cdot B\bm{\Omega}+\bm{\Lambda}\cdot \bm{U}$, we have 
$\displaystyle\fdv{\mathcal{E}}{\psi_s}
=
\left(
\begin{matrix}
    \widehat{B\bm{\Omega}} & \bm{\Lambda}\\
    0 & 0
\end{matrix}
\right)
$, 
and the variation of $E=\displaystyle\int_{-L/2}^{L/2}\mathcal{E}\dd{s}$ generated by $\psi_\xi$ is
\begin{gather}
\label{eq:delta_E}
    \delta E = \int_{-L/2}^{L/2}\left\langle\fdv{\mathcal{E}}{\psi_s},\var{\psi_s}\right\rangle\dd{s}
    =
    \left.\left\langle\fdv{\mathcal{E}}{\psi_s},\psi_\xi\right\rangle\right|_{-L/2}^{L/2}
    +
    \int_{-L/2}^{L/2}\left\langle-\partial_s\fdv{\mathcal{E}}{\psi_s}+\coad{\psi_s}\fdv{\mathcal{E}}{\psi_s},\psi_\xi\right\rangle\dd{s}.
\end{gather}
Expressing active and viscous stress on the body 
by the $\mathfrak{se}(3)$-valued matrix 
$
\mathcal{N}
=
\left(
\begin{matrix}
    -\zeta_r\bhat{\omega}_\parallel+m \bhat{U} &  -\zeta \bm{u}+f \bm{U}
    \\
    0&0
\end{matrix}
\right)
$, the virtual work $\mathcal{W}=\displaystyle\int_{-L/2}^{L/2}\left\langle\mathcal{N},\psi_\xi\right\rangle\dd{s}$ done by $\mathcal{N}$, as a result of the variation $\psi_\xi$, balances \eqref{eq:delta_E}, $\var{E}=\mathcal{W}$, to give the Euler-Poincar{\'e} equations,
\begin{gather}
\label{eq:ep_equations}
    \partial_s\fdv{\mathcal{E}}{\psi_s}-\coad{\psi_s}\fdv{\mathcal{E}}{\psi_s} = -\mathcal{N}.
\end{gather}
Separating \eqref{eq:ep_equations} into its linear and angular parts gives the force and moment balance equations found in the main text. The natural boundary conditions, given by $ \left.\displaystyle\fdv{\mathcal{E}}{\psi_s}\right|_{s=\pm L/2}=0$, are discussed below.

\subsection{Derivation of the shape evolution equations}
\label{sec:shape_evolution_equations}
The shape evolution is completely described by five vector equations equations:
\begin{gather}
    \label{eq:moment_balance}
    \partial_s\left(B\bm{\Omega}\right)+\bm{\Omega}\times\left(B\bm{\Omega}\right)+\bm{U}\times\bm{\Lambda} +m\bm{U}-\zeta_r\bm{\omega}_\parallel=\bm{0},
    \\
    \label{eq:force_balance}
    \partial_s\bm{\Lambda}+\bm{\Omega}\times\bm{\Lambda} + f\bm{U}-\zeta\bm{u}=\bm{0},
    \\
    \label{eq:angular_structure}
    \partial_s\bm{\omega}-\partial_t\bm{\Omega} + \bm{\Omega}\times\bm{\omega}=\bm{0},
    \\
    \label{eq:linear_structure}
    \partial_s\bm{u}-\partial_t\bm{U} + \bm{\Omega}\times\bm{u}-\bm{\omega}\times\bm{U}=\bm{0},
    \\
    \label{eq:U_constraint}
    \bm{U}=\bm{e}_0.
\end{gather}
Eqs.~\eqref{eq:moment_balance}-\eqref{eq:force_balance}, Eq.~\eqref{eq:U_constraint}, and the transverse part of Eq.~\eqref{eq:linear_structure} are used to express $(\bm{\omega},\bm{u}, \bm{\Lambda}_\perp)$,
\begin{gather}
\label{eq:lambda_perp}
\bm{\Lambda}_\perp
=
\bm{U}\times\partial_s\left(B_\perp\bm{\Omega}_\perp\right)
+
\left(B_\parallel-B_\perp\right)
\left(\bm{U}\cdot\bm{\Omega}_\parallel\right)
\bm{\Omega}_\perp,
\\
\label{eq:u}
\bm{u}
= \zeta^{-1}\left(\partial_s\bm{\Lambda}+\bm{\Omega}\times\bm{\Lambda} + f\bm{U}\right),
\\
\label{eq:w}
\bm{\omega}
= \frac{1}{\zeta_r}\left(\partial_s\left(B_\parallel\bm{\Omega}_\parallel\right) + m\bm{U}\right)
+
\bm{U}\times\left(\partial_s\bm{u}+\bm{\Omega}\times\bm{u}\right),
\end{gather}
in terms of $(\bm{\Omega},\bm{\Lambda}_\parallel, m,f)$ and their spatial derivatives. 
Eq.~\eqref{eq:angular_structure} and the remaining $\bm{U}$-component of \eqref{eq:linear_structure},
\begin{gather}
    \label{eq:angular_structure_parallel}
    \partial_t\bm{\Omega}_\parallel = \partial_s\bm{\omega}_\parallel
    + \bm{\Omega}_\perp\times\bm{\omega}_\perp,
    \\
    \label{eq:angular_structure_perp}
    \partial_t\bm{\Omega}_\perp = \partial_s\bm{\omega}_\perp
    + \bm{\Omega}_\parallel\times\bm{\omega}_\perp
    + \bm{\Omega}_\perp\times\bm{\omega}_\parallel,
    \\
     \label{eq:linear_structure_parallel}
    \partial_s\bm{u}_\parallel + \bm{\Omega}_\perp\times\bm{u}_\perp=\bm{0},
\end{gather}
 with $\bm{\omega}=\bm{\omega}[\bm{\Omega},\bm{\Lambda}_\parallel, m,f]$, $\bm{u}=\bm{u}[\bm{\Omega},\bm{\Lambda}_\parallel, f]$, and $\bm{\Lambda}_\perp=\bm{\Lambda}_\perp[\bm{\Omega}]$,
govern shape evolution and the tension. In terms of $(\bm{\Omega},\lambda,f)$, the shape evolution equations \eqref{eq:angular_structure_parallel}, \eqref{eq:angular_structure_perp}, \eqref{eq:linear_structure_parallel}, and the active force evolution equation, take the form
\begin{gather}
\label{eq:shape_evolution}
\begin{aligned}
    \partial_t\bm{\Omega}_\parallel
    -
    \dot{\bm{\Omega}}_\parallel\left(\bm{\Omega}_\parallel,\partial_s \bm{\Omega}_\parallel, \partial_s^2 \bm{\Omega}_\parallel, \vb{\Omega}_\perp,\ldots,\partial_s^3 \vb{\Omega}_\perp, \partial_s f, \lambda\right)
    =
    \bm{0},
    \\
    \partial_t \bm{\Omega}_\perp
    -
    \dot{\bm{\Omega}}_\perp
    \left(\bm{\Omega}_\parallel,\ldots,\partial_s^3 \bm{\Omega}_\parallel, \vb{\Omega}_\perp,\ldots,\partial_s^4 \vb{\Omega}_\perp, f, \partial_s f, \lambda, \ldots, \partial_s^2 \lambda \right)
    =
    \bm{0},
    \\
    \partial_s^2\lambda-\frac{1}{\eta}|\bm{\Omega}_\perp|^2\lambda
    +
    F_\lambda\left(\bm{\Omega}_\parallel, \vb{\Omega}_\perp,\ldots,\partial_s^2 \vb{\Omega}_\perp, \partial_s f\right)
    =
    0,
    \\
    \partial_t f - \dot{f}(\vb{\Omega}_\perp,\partial_s\vb{\Omega}_\perp, f, \partial_s^2 f, \partial_s\lambda)
    =
    0,
\end{aligned}
\end{gather}
where $\dot{\bm{\Omega}}_\parallel$, $\dot{\bm{\Omega}}_\perp$, $F_\lambda$, and $\dot{f}$ are nonlinear functions. Natural boundary conditions obtained from \eqref{eq:delta_E} state $\bm{\Lambda}=0$ and $\bm{\Omega}=0$ at $s=\pm L/2$. Hence, using \eqref{eq:lambda_perp}, we obtain
\begin{gather}
\label{eq:shape_bcs}
\bm{\Omega}_\parallel(s=\pm L/2)=\bm{0},\,\, \lambda(\pm L/2) = 0,\,\, \bm{\Omega}_\perp(\pm L/2)=\bm{0},\,\, \partial_s\bm{\Omega}_\perp(\pm L/2)=\bm{0}.
\end{gather}

Writing $\kappa^2=|\vb{\Omega}_\perp|^2$, the tension equation is given by

\begin{gather}
\label{eq:tension}
\begin{aligned}
&
\partial_{s}^2\lambda-\frac{\zeta_\parallel}{\zeta_\perp}\kappa^2\lambda+\frac{B_\perp}{2}\partial_{s}^2\kappa^2+\partial_{s}f\\
&
  +
  \frac{\zeta_\parallel B_\perp}{\zeta_\perp}
  \left(
  \vb{\Omega}_\perp\cdot\partial_{s}^2\vb{\Omega}_\perp-\kappa^2|\vb{\Omega}_\parallel|^2-2\vb{\Omega}_\parallel\cdot\left[\vb{\Omega}_\perp\times\partial_{s}\vb{\Omega}_\perp\right]
  \right)\\
  &
  +
  \frac{\zeta_\parallel B_\parallel}{\zeta_\perp}
  \left(
  \kappa^2|\vb{\Omega}_\parallel|^2+\vb{\Omega}_\parallel\cdot\left[\vb{\Omega}_\perp\times\partial_{s}\vb{\Omega}_\perp\right]
  \right)
  =0.
  \end{aligned}
\end{gather}

\section{Numerics}
\label{sec:numerics}
\subsection{Shape evolution}
We choose a uniform spatial grid, $s_n = n\Ds$ with $\Ds=L/2N$ for $n=-(N+2),\ldots, N+2$ and approximate spatial derivatives using second-order central differences. The introduction of ghost points $s_{-(N+2)}=-L-2\Ds,$ $s_{-(N+1)}=-L-\Ds,$ $s_{N+1}=L+\Ds,$ and $s_{N+2}=L+2\Ds$ allow central differences to be used at all interior and boundary points $s_{-N}=-L/2,...,s_N=L/2$. Time derivatives are approximated using second-order backwards differentiation for $t>\Dt$, and first order backward Euler for initialization, e.g.
\begin{gather}
\partial_t f(t) = \frac{3f(t)-4f(t-\Dt)+f(t-2\Dt)}{2\Dt}+O(\Dt^2),
    \\
    \partial_t f(t) = \frac{f(t)-f(t-\Dt)}{\Dt}+O(\Dt).
\end{gather}

Let $X_t=[\ldots \bm{\Omega}(s_n,t)\ldots f(s_n,t)\ldots\lambda(s_n,t)\ldots]^T$ denote the vector of samples at each spatial gridpoint at time $t$, and let $G=G(X_{t-2\Dt}, X_{t-\Dt}, X_{t})$ be the discretization of \eqref{eq:shape_evolution}-\eqref{eq:shape_bcs}. The function $(X_{t-2\Dt}, X_{t-\Dt}, X_{t})\mapsto G(X_{t-2\Dt}, X_{t-\Dt}, X_{t})$ was constructed using the symbolic computational library SymPy. At each timestep we compute $X_t$ from $(X_{t-2\Dt}, X_{t-\Dt})$ by applying Scipy's scipy.optimize.root to the function $X_t\mapsto G(X_{t-2\Dt}, X_{t-\Dt}, X_{t})$.

\subsection{Centerline and material frame evolution}
Time evolution of $\rho$ is performed in two steps. First, we evolve the $s=0$ cross-section in time using
\begin{gather}
    \label{eq:rho0_evo_exact}
    \rho(0, t) = \rho(0,t-\Dt)\Pexp_{0}^{\Dt}\psi_t(0,t-\Dt+\xi)\dd{\xi}.
\end{gather}
Next, the position and orientation of each $s\neq 0$ cross-section is integrated using
\begin{gather}
    \rho(s, t) = \rho(0,t)\Pexp_{0}^{\Ds}\psi_s(\xi,t)\dd{\xi}
    \Pexp_{0}^{\Ds}\psi_s(\Ds+\xi,t)\dd{\xi}
    \cdots
    \Pexp_{0}^{\Ds}\psi_s(s-\Ds+\xi,t)\dd{\xi}.
\end{gather}
We approximate ordered exponentials using explicit second order accurate Magnus integrators, guaranteeing $\rho(s,t)\in SE(3)$. With $\phi(\xi):=\psi_t(0, t-\Dt+\xi)$, the evolution operator, $\Pexp_{0}^{\xi}\phi(\sigma)\dd{\sigma}=e^{\theta(\xi)}$, can be expressed in terms of $\theta(\xi)\in \mathfrak{se}(3)$ which is computed by Magnus expansion
\begin{gather}
\label{eq:magnus}
    \theta(\xi)
    =
    \int_0^\xi\phi(\sigma)\dd{\sigma}-\frac{1}{2}\int_0^\xi\int_0^{\sigma}\comm{\phi(\sigma')}{\phi(\sigma)}\dd{\sigma'}\dd{\sigma}
    +
    \ldots
\end{gather}
Numerically integrating the first term using the trapezoid rule is sufficient for second order accuracy, 
\begin{gather}
    \theta(\xi) = \xi\frac{\phi(\xi)+\phi(0)}{2} + O(\xi^3).
\end{gather}

At each timestep, after computing $(X_{t-\Dt},X_{t})$, we use \eqref{eq:u}-\eqref{eq:w} to compute $(\psi_t(0,t-\Dt), \psi_t(0,t))$ in terms of $(X_{t-\Dt},X_{t})$ and their spatial differences. The $s=0$ cross-section is computed to second order in $\Dt$,
\begin{gather}
\label{eq:rho0_update}
    \rho(0, t) = \rho(0, t-\Dt)\exp(\Dt\frac{\psi_t(0, t)+\psi_t(0, t-\Dt)}{2})+O(\Dt^3),
\end{gather}
with $\exp$ given by \eqref{eq:se3_exp}. Similarly, spatial integration of $\rho$ gives
\begin{gather}
\label{eq:rhos_update}
    \rho(s, t) = \rho(0, t)\ldots\exp(\Ds\frac{\psi_s(s-\Delta s, t)+\psi_s(s-2\Ds, t)}{2})\exp(\Ds\frac{\psi_s(s, t)+\psi_s(s-\Ds, t)}{2}) +O(\Ds^3).
\end{gather}
Exponentials are computed using the closed form expression for the $SE(3)$ exponential map found above.

\section{2D dynamics ($M=0$)}
In the $M=0$ case, $\bm{\Omega} = [0,0,\kappa]^T$ has a single nonzero component given by the signed centerline curvature, $\kappa$. Writing $\beta:=\beta_\perp$, we have 
\begin{gather}
    \partial_{t}\kappa
    =
    -\frac{\beta}{\eta}\partial_{s}^{4}\kappa
+
\frac{\beta}{3}\partial_s^{2}\left(
 \kappa^{3}
\right)
+\frac{1}{\eta}\partial_{s}^{2}\left(
\kappa \lambda
\right)
+
\partial_s\left(
\kappa\left[
\partial_s\lambda+f
\right]
\right),
\\
   \partial_{s}^{2}\lambda
   - \frac{\kappa^{2}}{\eta}\lambda
   =
   - \partial_{s}f
-
\frac{\beta}{2}\partial_s^2
\left(
 \kappa^2
\right)
-
\frac{\beta }{\eta}\kappa \partial_{s}^{2}\kappa,
\\
\partial_{t}f
   =
    \sigma\partial_{s}^{2}f  + \left(1 - f^{2}\right) \left(\beta \kappa \partial_{s}\kappa + \partial_{s}\lambda + f\right),
\end{gather}
with boundary conditions $\kappa=0,\partial_s\kappa=0,\lambda,\partial_s f$ at $s=\pm L/2$

\section{Linearization}
\label{sec:2d_linearization}
To first order in $\kappa$, with $\overline{f}:=\displaystyle\int_{-1/2}^{1/2}f\dd{s}$, we find

\begin{gather}
    \lambda =- \int_{-1/2}^{s}(f-\overline{f})\dd{s},
    \\
    \partial_{t}\kappa
=
-\frac{\beta}{\eta}\partial_{s}^{4}\kappa
-\frac{1}{\eta}\partial_{s}^{2}\left(\kappa\int_{-1/2}^{s}f-\overline{f}\dd{s}
\right)
+
\overline{f}\partial_s \kappa.
\end{gather}

When \(\overline{f}=0\), this reduces to
\begin{gather}
   \lambda =-  \int_{-1/2}^{s}f\dd{s},
   \\
   \partial_{t}\kappa
=
-\frac{\beta}{\eta}
\partial_{s}^{2}
\left(\partial_{s}^{2}\kappa
-\frac{1}{\beta}\lambda\kappa
\right),
 \end{gather}
 giving the eigenvalue problem,
\begin{gather}
\label{eq:eigen_problem}
\begin{aligned}
\mathcal{L}[\kappa]
=
-\frac{\beta}{\eta}
\partial_{s}^{2}
&
\left(\partial_{s}^{2}\kappa
-\frac{1}{\beta}\lambda\kappa
\right),
\\
\mathcal{L}[\kappa]&=\mu\kappa.
\end{aligned}
 \end{gather}

\subsection{Critical $\beta$ values for $f(s) = 1-2H(s)$}
With the piecewise constant active force, $f(s) = 1-2H(s)$, with $H(s)$ the Heaviside step function, the tension is given by
\begin{gather}
    \lambda(s) = 2sH(s)-(s+1/2)
    =
    \begin{cases}
        -(s+1/2) &  -1/2<s<0,\\
        s-1/2 &  0<s<1/2.
    \end{cases}
\end{gather}
Critical values of $\beta$ are associated with the emergence of nontrivial solutions to~\ref{eq:eigen_problem} with $\mu=0$, which, after integrating twice is equivalent to
\begin{gather}
\partial_s^2\kappa - \frac{1}{\beta}\lambda\kappa = c^{(-)} s+c^{(+)}, 
\end{gather}
for constants $(c^{(-)}, c^{(+)})$. When $\lambda(s)=\lambda(-s)$, Eq.~\eqref{eq:eigen_problem} is invariant under $\kappa(s)\mapsto\kappa(-s)$, so we may assume eigenfunctions have definite parity. When $\kappa(-s)=\kappa(s)$, we find $c^{(-)}=0$, and when $\kappa(-s)=-\kappa(s)$ we find $c^{(+)}=0$. Restricting to the half interval, $-1/2<s<0$ and introducing $\xi=\beta^{-1/3}\lambda=-(1+2s)/\left(2\beta^{1/3}\right)$, we find $\kappa$ satisfies a nonhomogeneous Airy's equation on $-1/\left(2\beta^{1/3}\right)<\xi<0$,
\begin{gather}
\label{eq:airy_eq}
   \partial_\xi^2\kappa-\xi\kappa=a\xi+b,
   \\
   \label{eq:airy_bc}
   \left.\kappa\right|_{\xi=0},\left.\partial_\xi\kappa\right|_{\xi=0}=0,
\end{gather}
where $(a,b)=(-\beta^{1/3}c^{(-)}, c^{(+)}-c^{(-)}/2)$. The solution to~\eqref{eq:airy_eq}-\eqref{eq:airy_bc} is given by 
\begin{gather}
    \kappa(\xi) = -\pi\Ai(\xi)\int_0^{\xi}\left(ax+b\right)\Bi(x)\dd{x}+\pi\Bi(\xi)\int_0^{\xi}\left(ax+b\right)\Ai(x)\dd{x}\nonumber
    \\
    = \pi a\left[\Bi'(0)\Ai(\xi)-\Ai'(0)\Bi(\xi)-1/\pi\right]
    +\pi b\left[
    \Bi(\xi)\int_0^{\xi}\Ai(x)\dd{x} - \Ai(\xi)\int_0^{\xi}\Bi(x)\dd{x}
    \right],
\end{gather}
where $\Ai(\xi) = \displaystyle\frac{1}{\pi}\int_0^\infty \cos(x^3/3+\xi x)\dd{x}$ and $\Bi(\xi) = \displaystyle\frac{1}{\pi}\int_0^\infty \sin(x^3/3+\xi x) + e^{-x^3/3+\xi x}\dd{x}$ are Airy functions, and primes denote derivatives. Writing $\xi^*=-\beta^{-1/3}/2$, parity conditions require $(a, \left.\partial_\xi\kappa\right|_{\xi^*})=(0,0)$, giving
\begin{gather}
    \kappa(\xi) = \Bi(\xi)\int_0^{\xi}\Ai(x)\dd{x}-\Ai(\xi)\int_0^{\xi}\Bi(x)\dd{x},
    \\
     \Bi'(\xi^*)\int_0^{\xi^*}\Ai(x)\dd{x}-\Ai'(\xi^*)\int_0^{\xi^*}\Bi(x)\dd{x}=0,
\end{gather}
and $(b/a, \left.\kappa\right|_{\xi^*})=(-\xi^*,0)$, giving
\begin{gather}
    \kappa(\xi) =
    \Bi'(0)\Ai(\xi)-\Ai'(0)\Bi(\xi) -1/\pi
     -\xi^*\left(\Bi(\xi)\int_0^{\xi}\Ai(x)\dd{x} - \Ai(\xi)\int_0^{\xi}\Bi(x)\dd{x}\right)\nonumber
    \\
     \Bi'(0)\Ai(\xi^*)-\Ai'(0)\Bi(\xi^*) -1/\pi
     -\xi^*\left(\Bi(\xi^*)\int_0^{\xi^*}\Ai(x)\dd{x} - \Ai(\xi^*)\int_0^{\xi^*}\Bi(x)\dd{x}\right)
     =
     0,
\end{gather}
for even and odd eigenfunctions, respectively.

\subsection{Active force-moment ratio}
An approximation of the force/moment ratio for each flagellum on the body surface is needed for the continuum surface traction/moment model. The normal helical form of a peritrichous flagellum is left handed, with a helical pitch $P$ of roughly $2 \mu$m and helical circumference $C=1.5\mu$m, resulting in a pitch angle $\psi=\tan^{-1}(C/|P|) \approx 35^\circ$. Flagellar lengths are on the order of $\ell = 10\mu$m for organisms like {\it E. coli}. Flagella are all roughly $20$nm in diameter, $d$.

Consider such a helical body, its basal end fixed to a rigid wall, rotated about its long axis with speed $\omega$, generates an axial force on that wall (ignoring finite length effects) of magnitude 
\begin{gather}
F = \frac{2\mu C^2 L \omega}{P c},
\end{gather}
where $\mu$ is the viscosity of water and $c=\ln(L^2/d^2)-1$. Meanwhile, the viscous torque experienced by the helical body is given by 
\begin{gather}
L = \frac{2 \mu C^2 L \omega}{\pi^2 c}.
\end{gather}
Hence, $L/F = P/\pi^2 \approx 10^{-7}$m.

\section{Biharmonic equation }
\label{sec:biharmonic}
Unnormalized solutions to the biharmonic eigenvalue problem,
\begin{gather}
\phi:[-1/2,1/2]\rightarrow \mathbb{R},
\\
  \partial_s^4\phi_k = k^4\phi_k,
  \\
  \phi_{k}(\pm 1/2)=\partial_{s}\phi_{k}(\pm 1/2)=0,
\end{gather}
are given by
\begin{gather}
  \Phi_k = \left[\sinh(k)-\sin(k)\right]\left[\cos(k[s+1/2])-\cosh(k[s+1/2])\right]+\left[\cos(k)-\cosh(k)\right]\left[\sin(k[s+1/2])-\sinh(k[s+1/2])\right],
\end{gather}
where the eigenvalues $k^4$ must satisfy $ \cos(k)\cosh(k)=1$.
\subsection{Even/odd eigenfunctions}
Unnormalized even eigenfunctions, $\Phi^{+}_k$, must satisfy
\begin{gather}
  \Phi^{+}_k = \cosh(k/2) \cos(ks)-\cos(k/2) \cosh(ks),
  \\
  \tan(k/2)=-\tanh(k/2),
  \\
  \norm{ \Phi^{+}_k}^2
  =
  \int_{-1/2}^{1/2} |\Phi^{+}_k|^2\dd{s},
  =
  \frac{1}{2}\left[\cosh[2](k/2)+\cos[2](k/2)\right],
\end{gather}

while the unnormalized odd eigenfunctions, $\Phi^{-}_k$, satisfy
\begin{gather}
  \Phi^{-}_k = -\sinh(k/2) \sin(ks)+\sin(k/2) \sinh(ks),
  \\
  \tan(k/2)=\tanh(k/2),
  \\
   \norm{ \Phi^{-}_k}^2,
   =
   \int_{-1/2}^{1/2} |\Phi^{-}_k|^2\dd{s}
   =
   \frac{1}{2}\left[\sinh[2](k/2)-\sin[2](k/2)\right],
\end{gather}

The following normalized even, $\phi_k^{+}:=\Phi^{+}_k/\norm{\Phi^{+}_k}$, and odd, $\phi_k^{-}:=\Phi^{-}_k/\norm{\Phi^{-}_k}$, functions were used in numerical calculations:
\begin{gather}
    \phi_k^{+} = \frac{\sqrt{2}\cos(ks)}{\sqrt{1+\displaystyle\frac{\cos[2](k/2)}{\cosh[2](k/2)}}}
  -
  \frac{\sqrt{2}\cos(k/2)}{(1+e^{-k})\sqrt{1+\displaystyle\frac{\cos[2](k/2)}{\cosh[2](k/2)}}}
  \left(e^{-k(1/2-s)}+e^{-k(s-1/2)}\right),
\end{gather}

\begin{gather}
    \phi_k^{-}
    =
    \displaystyle\frac{-\sqrt{2}\sin(ks)}{\sqrt{1-\frac{\sin[2](k/2)}{\sinh[2](k/2)}}}
  +
  \frac{\sqrt{2}\sin(k/2)}{(1-e^{-k})\sqrt{1-\displaystyle\frac{\sin[2](k/2)}{\sinh[2](k/2)}}}\left(e^{-k(1/2-s)}-e^{-k(s+1/2)}\right).
\end{gather}

\bibliographystyle{unsrt}
\bibliography{BigBib}